\documentclass[12pt]{iopart}
\usepackage{graphicx}
\usepackage{indentfirst}

\begin{document}

\title[]{Enhancement of non-equilibrium thermal quantum discord and entanglement of a
three-spin XX chain by multi-spin interaction and external magnetic
field}
\author{Xiu-xing Zhang and Fu-li Li$^{1}$}
\address{MOE Key Laboratory for Nonequilibrium Synthesis and
Modulation of Condensed Matter, and Department of Applied Physics,
Xi'an Jiaotong University, Xi'an 710049, China}
\ead{flli@mail.xjtu.edu.cn}

\begin{abstract}
We investigate the non-equilibrium thermal quantum discord and
entanglement of a three-spin chain whose two end spins are
respectively coupled to two thermal reservoirs at different
temperatures. In the three-spin chain, besides the XX-type
nearest-neighbor two-spin interaction, a multi-spin interaction is
also considered and a homogenous magnetic field is applied to each
spin. We show that the extreme steady-state quantum discord and
entanglement of the two end spins can always be created by holding
both a large magnetic field and a strong multi-spin interaction. The
results are explained by the thermal excitation depression due to
switching a large energy gap between the ground state and the first
excited state. The present investigation may provide a useful
approach to control coupling between a quantum system and its
environment.
\end{abstract}

\pacs{03.65.Yz, 03.67.Mn, 03.67.Bg}
$^1$ Author to whom any
correspondence should be addressed. \maketitle

\newpage

\section{Introduction}

Quantum entanglement was once considered as a unique resource that can be
used in quantum information processing \cite{Nielsen}. However, recent
researches have shown that besides entanglement a composite quantum system
may have other kinds of nonclassical correlations which can appear even in
separable states \cite{Henderson,Vedral,ex1,Knill,Ollivier}. In order to
quantitatively describe such quantum correlations in a composite quantum
system, many different measures have been proposed \cite%
{Ollivier,measure2,measure4,measure5}. Among them, quantum discord (QD),
firstly introduced by Ollivier and Zurk \cite{Ollivier}, has received
considerable attention \cite%
{analytical,QPT1,QPT2,QPT3,QPT4,variable1,variable2}. Using an optical
architecture, Lanyon \textit{et al. }\cite{ex1} experimentally showed that
even fully separable states with quantum discord can be used to construct
quantum computer.

In all real situations, quantum systems can not be completely isolated from
their environments. Coupling of a quantum system to its surrounding
unavoidablely results in the destruction of quantum correlations of the
system. The effect of environments to entanglement of bipartite quantum
systems have intensively been investigated. It has been shown that
entanglement undergoes sudden death due to the interaction of quantum
systems with their reservoirs \cite{ESD}. In recent years, the QD dynamics
of open quantum systems has also attracted much interest in both theory and
experiment \cite{Markovian, non-Markovian1, non-Markovian2, sudden tran 1,
sudden tran 2, Quantum dot, Main Ref,ex2,ex3,ex4}. Werlang \textit{et al. }
\cite{Markovian} investigated the dynamics of both entanglement and QD in
the Markovian environments, and showed that QD is more robust against
decoherence than entanglement. Recently, an interesting dynamical feature of
QD, named sudden transition, has been observed \cite{sudden tran 1, sudden
tran 2}. It means that for certain initial states QD undergoes sudden change
between a \textquotedblleft classical decoherence" phase and a
\textquotedblleft quantum decoherence" phase \cite{non-Markovian2, sudden
tran 1}. This sudden transition behavior can be explained in a geometrical
way and has connection with the property of environment \cite%
{non-Markovian2, sudden tran 1}. Xu \textit{et al. }\cite{ex2, ex3}
experimentally investigated both the Markovian and non-Markovian dynamics of
classical and quantum correlations and observed the sudden transition
behavior of QD.

Apart from the situation in which a quantum system is coupled to a
single environment, it may also be possible that a quantum system is
simultaneously in contact with two different thermal baths. In
semiconductor quantum dots nuclear spins and electronic spins
consist of a composite quantum system for quantum information
processing and quantum computing but the coupling manners of nuclear
spins and electron spins to their surroundings are much different
\cite{env1,env2,env3}. With the help of NMR and quantum optical
techniques, one can create two reservoirs at different effective
temperatures for nuclear spins and electron spins in quantum dots
\cite {nonex1,nonex2}. For superconductor qubits, the
two-different-thermal-bath coupling situation may directly be
designed \cite{qif}. When interacting with two reservoirs at
different temperatures, a quantum system may approach a steady state
instead of a thermal equilibrium state. Thus, in general, the
presence of heat/energy/mass currents passing through the quantum
system in a steady-state may modify the quantum correlations.

In recent years, quantum correlations of coupled qubits in contact
with two different thermal environments, i.e., the non-equilibrium
thermal environment model, have received some attention
\cite{non1,non2,non3,non4,Yi}. Quiroga \textit{et al. }\cite{non3}
proposed a two-interacting spin-$1/2$ system in contact with two
heat reservoirs at different temperature, and identified a
nonequilibrium enhancement-suppression transition behavior of
entanglement due to the presence of temperature gradient. Employing
the same model, Sinaysky \textit{et al. }\cite{non4} found that the
spin system can converge to steady state and studied the dependence
of the steady-state concurrence on the mean temperature and
temperature difference of the reservoirs and the energy splits of
the spins. Huang \textit{et al. }\cite {Yi} investigated the
nonequilibrium thermal steady-state entanglement in a three
spin-$1/2$ XX chain in contact with two heat reservoirs at different
temperature and found that the temperature difference of the heat
baths benefits the entanglement in the nonsymmetric coupling case.
Spin chain models in contact with two reservoirs at two different
temperatures have also been employed for studying heat current
transfer \cite{current1, current2,current3}. Yan \textit{et al.
}\cite{current1} considered an interacting spin-$1/2$ chain
connected to two phonon baths held at different temperatures and
showed that heat transport through the spins systems can be
controlled by applying an inhomogeneous magnetic field due to
switching an energy gap.

Stimulated by the pervious investigations, we here study how to
control the steady-state QD and entanglement of spin systems. We
consider a three-spin-$1/2$ chain in which besides the XX-type
nearest-neighbor two-spin interaction a three-spin interaction is
included and an external magnetic field is homogeneously applied to
each spin, and meanwhile the two end spins are coupled to two
thermal environments at different temperatures. We show that the
coupling of the spin system to the thermal reservoirs can be
controlled and the thermal excitation can be greatly depressed by
the three-spin interaction and the magnetic field. As a result, the
extreme steady-state QD and entanglement in the two end spins can be
created.

This paper is organized as follows. In Sec. II, we describe the model and
introduce the calculation method. In Sec. III, definitions on quantum
discord and concurrence are briefly reviewed. In Sec. IV, numerical results,
discussion and physical explanations are presented. Finally, a summary is
given in Sec. V.

\section{Model And Master Equation}

\begin{figure}[tbph]
\centering \includegraphics[width=12.0cm]{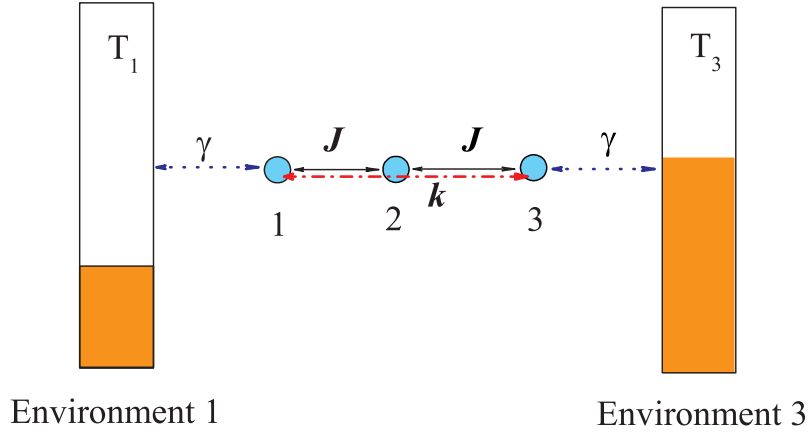} \caption{A
schematic representation of a three-spin chain coupled to two
thermal baths at different temperatures, $T_{1}$ and $T_{3}$.}
\label{fig:Fig1}
\end{figure}
The model under investigation is described in Fig. 1. We consider a
three-spin-$1/2$ chain which Hamiltonian reads
\begin{equation}
H_{S}=J\sum_{i=1}^{2}\left( \sigma _{i}^{x}\sigma _{i+1}^{x}+\sigma
_{i}^{y}\sigma _{i+1}^{y}\right)+h\sum_{i=1}^{3}\sigma _{i}^{z} +k\left(
\sigma _{1}^{x}\sigma _{2}^{z}\sigma _{3}^{x}+\sigma _{1}^{y}\sigma
_{2}^{z}\sigma _{3}^{y}\right) ,  \label{Hamiltonian}
\end{equation}
where $\sigma _{i}^{\alpha }(\alpha =x,y,z)$ are the Pauli matrices
for the $i$th spin, $J$ is the coupling constant between the
nearest-neighbor spins, and $h$ is the external magnetic field
strength, homogeneously applied to each spin. Besides the two-spin
interaction, the three-spin interaction \cite {multi1,multi2,multi3}
is also included, which strength is denoted by $k$.

As shown in Fig. 1, two end spins $1$ and $3$ are in contact with
two phonon baths at different temperatures, $T_{1}$ and $T_{3} $,
respectively. In the interaction picture, the Hamiltonian describing
the interaction between the $j$th spin and its phonon bath is given
by
\begin{equation}
H_{SBj}=\sigma _{j}^{x}\left( \sum_{n}g_{j}^{(n)}e^{-i\omega
_{nj}t}b_{nj}+g_{j}^{(n)\ast }e^{i\omega _{nj}t}b_{nj}^{\dagger }\right)
\equiv\sigma _{j}^{x}\otimes B_{j},(j=1,3),  \label{inter}
\end{equation}%
where $b_{nj}^{\dagger }(b_{nj})$ is the creation (annihilation) operator
for the $n$th mode of thermal bath $j$, and $g_{j}^{(n)}$ is the coupling
constant between the $j$th spin and the $n$th bath mode.

Let us first consider the eigenvalue problem of the Hamiltonian (1)
\begin{equation}
H_{S}\left\vert \phi_l \right\rangle =\varepsilon_l \left\vert \phi_l
\right\rangle, (l=1,2,3,...,8) .  \label{eigen equation}
\end{equation}%
The spin-up and spin-down states of spin $i$ are represented by
state-vectors $\left\vert 1\right\rangle _{i}$ and $\left\vert
0\right\rangle _{i}$, respectively. In the presentation spanned by the
uncoupled basis $\left\vert n_{1}n_{2}n_{3}\right\rangle =\left\vert
n_{1}\right\rangle _{1}\otimes \left\vert n_{2}\right\rangle _{2}\otimes
\left\vert n_{3}\right\rangle _{3}$ with $n_{i}=0,1$, we can easily work out
the eigenstates of Eq. (\ref{eigen equation}) as follows
\begin{eqnarray}
\left\vert \phi _{1}\right\rangle &=&\left\vert 000\right\rangle ,
\label{EV1} \\
\left\vert \phi _{2}\right\rangle &=&\left\vert 111\right\rangle ,
\label{EV2} \\
\left\vert \phi _{3}\right\rangle &=&\frac{1}{\sqrt{2}}\left( -\left\vert
110\right\rangle +\left\vert 011\right\rangle \right) ,  \label{EV3} \\
\left\vert \phi _{4}\right\rangle &=&\frac{1}{\sqrt{2}}\left( -\left\vert
100\right\rangle +\left\vert 001\right\rangle \right) ,  \label{EV4} \\
\left\vert \phi _{5}\right\rangle &=&\frac{1}{\sqrt{2}}\sin \alpha
_{1}\left\vert 100\right\rangle +\cos \alpha _{1}\left\vert 010\right\rangle
+\frac{1}{\sqrt{2}}\sin \alpha _{1}\left\vert 001\right\rangle ,  \label{EV5}
\end{eqnarray}
\begin{eqnarray}
\left\vert \phi _{6}\right\rangle &=&\frac{1}{\sqrt{2}}\sin \alpha
_{2}\left\vert 110\right\rangle -\cos \alpha _{2}\left\vert 101\right\rangle
+\frac{1}{\sqrt{2}}\sin \alpha _{2}\left\vert 011\right\rangle ,  \label{EV6}
\\
\left\vert \phi _{7}\right\rangle &=&\frac{1}{\sqrt{2}}\sin \alpha
_{2}\left\vert 100\right\rangle +\cos \alpha _{2}\left\vert 010\right\rangle
+\frac{1}{\sqrt{2}}\sin \alpha _{2}\left\vert 001\right\rangle ,  \label{EV7}
\\
\left\vert \phi _{8}\right\rangle &=&\frac{1}{\sqrt{2}}\sin \alpha
_{1}\left\vert 110\right\rangle -\cos \alpha _{1}\left\vert 101\right\rangle
+\frac{1}{\sqrt{2}}\sin \alpha _{1}\left\vert 011\right\rangle ,  \label{EV8}
\end{eqnarray}
with the corresponding eigenvalues $\varepsilon _{1}=-3h,\varepsilon
_{2}=3h,\varepsilon _{3}=h-2k,\varepsilon _{4}=-h+2k,\varepsilon
_{5}=-h-k-B,\varepsilon _{6}=h+k-B,\varepsilon
_{7}=-h-k+B,\varepsilon _{8}=h+k+B$, where $B=\sqrt{8+k^{2}},\sin
\alpha _{1}=2\sqrt{2}/\sqrt{8+(k-B)^{2}}$, $\cos \alpha _{1}=\left(
k-B\right) /\sqrt{8+(k-B)^{2}}$, $\sin \alpha
_{2}=2\sqrt{2}/\sqrt{8+(k+B)^{2}}$ and $\cos \alpha _{2}=\left(
k+B\right) /\sqrt{8+(k+B)^{2}}$.

In the representation spanned by eigenstates (\ref{EV1})-(\ref{EV8}), the
Hamiltonian of the coupled reservoir -spin system can be written as
\begin{equation}
H=H_{S}+H_{SB1}+H_{SB3}=\sum_{l=1}^{8}\varepsilon _{l}\left\vert \phi
_{l}\right\rangle \langle \phi _{l}|+\sum_{j=1,3}\sum_{\omega }A_{j}(\omega
)\otimes B_{j},  \label{TH}
\end{equation}%
where
\begin{equation}
A_{j}(\omega )=\sum_{\varepsilon _{l}-\varepsilon _{l^{\prime }}=\omega
}\langle \phi _{l}|\sigma _{j}^{x}\left\vert \phi _{l^{\prime
}}\right\rangle \left\vert \phi _{l}\right\rangle \langle \phi _{l^{\prime
}}|.  \label{superoperator}
\end{equation}%
In Eq.(\ref{TH}), the summation $\sum_{\omega }$ must be done over
all possible differences between any two eigenenergies of the
Hamiltonian (1). In Eq. (\ref{superoperator}), the summation
$\sum_{\varepsilon _{l}-\varepsilon _{l^{\prime }}=\omega }$ is over
all the eigenvalues with a fixed difference $\omega $. Obviously,
$A^{\dagger }(\omega )=A(-\omega )$. Upon substitution of
eigenstates (\ref{EV1})-(\ref{EV8}) into Eq. (\ref{superoperator}),
we find the following nonzero transition operators
\begin{eqnarray}
A_{1}^{\dagger }\left( \omega _{1}\right)  &=&\frac{1}{\sqrt{2}}(\sin \alpha
_{1}\left\vert \phi _{2}\right\rangle \left\langle \phi _{8}\right\vert
-\cos \alpha _{2}\left\vert \phi _{6}\right\rangle \left\langle \phi
_{4}\right\vert   \nonumber \\
&&-\cos \alpha _{2}\left\vert \phi _{3}\right\rangle \left\langle \phi
_{7}\right\vert +\sin \alpha _{1}\left\vert \phi _{5}\right\rangle
\left\langle \phi _{1}\right\vert ),  \label{tt1}
\end{eqnarray}%
\begin{eqnarray}
A_{1}^{\dagger }\left( \omega _{2}\right)  &=&\frac{1}{\sqrt{2}}(\sin \alpha
_{2}\left\vert \phi _{2}\right\rangle \left\langle \phi _{6}\right\vert
-\cos \alpha _{1}\left\vert \phi _{8}\right\rangle \left\langle \phi
_{4}\right\vert   \nonumber \\
&&-\cos \alpha _{1}\left\vert \phi _{3}\right\rangle \left\langle \phi
_{5}\right\vert +\sin \alpha _{2}\left\vert \phi _{7}\right\rangle
\left\langle \phi _{1}\right\vert ),  \label{tt2}
\end{eqnarray}%
\begin{eqnarray}
A_{1}^{\dagger }\left( \omega _{3}\right)  &=&\frac{1}{\sqrt{2}}(\left\vert
\phi _{2}\right\rangle \left\langle \phi _{3}\right\vert -\sin \left( \alpha
_{-}\right) \left\vert \phi _{6}\right\rangle \left\langle \phi
_{5}\right\vert   \nonumber \\
&&+\sin \left( \alpha _{-}\right) \left\vert \phi _{8}\right\rangle
\left\langle \phi _{7}\right\vert -\left\vert \phi _{4}\right\rangle
\left\langle \phi _{1}\right\vert ),  \label{tt3}
\end{eqnarray}%
\begin{eqnarray}
A_{3}^{\dagger }\left( \omega _{1}\right)  &=&\frac{1}{\sqrt{2}}(\sin \alpha
_{1}\left\vert \phi _{2}\right\rangle \left\langle \phi _{8}\right\vert
+\cos \alpha _{2}\left\vert \phi _{6}\right\rangle \left\langle \phi
_{4}\right\vert   \nonumber \\
&&+\cos \alpha _{2}\left\vert \phi _{3}\right\rangle \left\langle \phi
_{7}\right\vert +\sin \alpha _{1}\left\vert \phi _{5}\right\rangle
\left\langle \phi _{1}\right\vert ),  \label{tt4}
\end{eqnarray}%
\begin{eqnarray}
A_{3}^{\dagger }\left( \omega _{2}\right)  &=&\frac{1}{\sqrt{2}}(\sin \alpha
_{2}\left\vert \phi _{2}\right\rangle \left\langle \phi _{6}\right\vert
+\cos \alpha _{1}\left\vert \phi _{8}\right\rangle \left\langle \phi
_{4}\right\vert   \nonumber \\
&&+\cos \alpha _{1}\left\vert \phi _{3}\right\rangle \left\langle \phi
_{5}\right\vert +\sin \alpha _{2}\left\vert \phi _{7}\right\rangle
\left\langle \phi _{1}\right\vert ),  \label{tt5}
\end{eqnarray}%
\begin{eqnarray}
A_{3}^{\dagger }\left( \omega _{3}\right)  &=&\frac{-1}{\sqrt{2}}(\left\vert
\phi _{2}\right\rangle \left\langle \phi _{3}\right\vert -\sin \left( \alpha
_{+}\right) \left\vert \phi _{6}\right\rangle \left\langle \phi
_{5}\right\vert   \nonumber \\
&&-\sin \left( \alpha _{+}\right) \left\vert \phi _{8}\right\rangle
\left\langle \phi _{7}\right\vert -\left\vert \phi _{4}\right\rangle
\left\langle \phi _{1}\right\vert ),  \label{tt6}
\end{eqnarray}%
where $\omega _{1}=2h-k-B,\omega _{2}=2h-k+B$, $\omega _{3}=2(h+k)$, $\alpha
_{+}=\alpha _{1}+\alpha _{2}$ and $\alpha _{-}=\alpha _{1}-\alpha _{2}$.

By means of the general reservoir theory within the Born-Markov and rotating
wave approximations \cite{master equation,open,Magn}, one can obtain the
equation of motion for the reduced density matrix of the spin chain
\begin{equation}
\frac{d\rho }{dt}=-i[H_{S},\rho ]+L_{1}(\rho )+L_{3}(\rho ),
\label{master equation}
\end{equation}%
where $L_{j}(\rho )$ ($j=1,3$) is the dissipative term due to the coupling
of spin $j$ to its thermal bath and is given by
\begin{eqnarray}
L_{j}(\rho ) &=&\sum_{\omega _{\mu }>0}\gamma _{j}(\omega _{\mu
})(1+n_{j}(\omega _{\mu }))\left( 2A_{j}\left( \omega _{\mu }\right) \rho
A_{j}^{\dagger }\left( \omega _{\mu }\right) -\left\{ \rho ,A_{j}^{\dagger
}\left( \omega _{\mu }\right) A_{j}\left( \omega _{\mu }\right) \right\}
\right)  \nonumber \\
&&+\sum_{\omega _{\mu }>0}\gamma _{j}(\omega _{\mu })n_{j}(\omega _{\mu
})\left( 2A_{j}^{\dagger }\left( \omega _{\mu }\right) \rho A_{j}\left(
\omega _{\mu }\right) -\left\{ \rho ,A_{j}\left( \omega _{\mu }\right)
A_{j}^{\dagger }\left( \omega _{\mu }\right) \right\} \right) .  \label{ld}
\end{eqnarray}

In deriving out the master equation (\ref{master equation}), we have
assumed that the $j$th bath is always in a thermal equilibrium state
at temperature $T_{j}$. In Eq. (\ref{ld}), $n_{j}(\omega _{\mu
})=1/(\exp (\beta _{j}\omega _{\mu })-1)$ with $\beta
_{j}=1/(T_{j})$ is the mean thermal photon number of the $j$th bath
at frequency $\omega _{\mu }$ (taking the Boltzmann constant
$k_{B}=1$), and $\gamma _{j}(\omega _{\mu })$ is defined through the
integral $\pi \sum_{n}|g_{j}^{(n)}|^{2}\left( 1+<b_{nj}^{\dagger
}b_{nj}>\right) =\int_{0}^{\infty }\gamma _{j}(\omega _{\mu
})(1+n_{j}(\omega _{\mu }))d\omega _{\mu }$. Here, the Lamb shift
has been omitted.

\section{Quantum Discord and Concurrence}

In this section, for convenience of discussions in the next section,
we give a brief review on quantum discord (QD) and concurrence. QD
is defined as the discrepancy between quantum extensions of two
equivalent expressions for the classical mutual information
\cite{Ollivier}. In classical information theory (CIT), the total
correlation between two random variables $A$ and $B$ can be
described by either the mutual information \cite{Nielsen, mutual1,
mutual2}
\begin{equation}
I_{C}\left( A\colon B\right) =H(A)+H(B)-H(A,B)  \label{CI1}
\end{equation}
or the equivalent expression
\begin{equation}
J_{C}\left( A\colon B\right) =H(A)-H(A\Vert B),  \label{CI2}
\end{equation}%
where $H(X)=-\sum_{x}p_{x}\log _{2}p_{x}$ $(X=A$, $B$ and $AB)$ is the
Shannon entropy of the variable $X$ with $p_{x}$ being the probability of $X$
assuming the value $x$, and $H(A\Vert B)=-\sum_{a,b}p_{ab}\log _{2}p_{a\mid
b}=H(A,B)-H(B)$ $(p_{a\mid b}=p_{ab}/p_{b})$ is the conditional entropy,
which represents a weighted average of the entropies of $A$ given the value
of $B$.

In the quantum information theory (QIT) \cite{Nielsen, mutual1, mutual2},
the total correlation of a bipartite system consisting of subsystems $A$ and
$B$ in a state described by the density matrix $\rho _{AB}$ is defined as
\begin{equation}
I_{q}\left( \rho _{A\colon B}\right) =S(\rho _{A})+S(\rho _{B})-S(\rho
_{AB}),  \label{TC}
\end{equation}%
which is the straightforward extension of (\ref{CI1}). Here, $S(\rho
_{A(B)})=-Tr(\rho _{A(B)}\log _{2}\rho _{A(B)})$ is the von Neumann entropy
of the subsystem $A(B)$, while $S(\rho _{AB})=-Tr(\rho _{AB}\log _{2}\rho
_{AB})$ is the entropy of the composite system $AB$.

The extension of (\ref{CI1}) to the quantum realm is no longer
straightforward since the value of $H(A\Vert B)$ is measurement dependence,
and quantum measurement may fully destroy a quantum state. The counterpart
of (\ref{CI2}) in QIT may be defined as
\begin{equation}
J_{q}\left( \rho _{A\colon B}\right) =S(\rho _{A})-S_{\left\{ \Pi
_{j}^{B}\right\} }\left( \rho _{A\mid B}\right) ,  \label{QI2}
\end{equation}%
where $\left\{ \Pi _{j}^{B}\right\} $\ are a set of projectors
performed locally on subsystem $B$, and $S_{\left\{ \Pi
_{j}^{B}\right\} }\left( \rho _{A\mid B}\right) =\sum_{j}q_{j}S(\rho
_{A}^{j})$ with $\rho _{A}^{j}=Tr_{B}\left( \Pi _{j}^{B}\rho
_{AB}\Pi _{j}^{B}\right) /q_{j}$ and the probability
$q_{j}=Tr_{AB}(\Pi _{j}^{B}\rho _{AB}\Pi _{j}^{B})$. The project
operator $\Pi _{j}^{B}=\left\vert \theta _{j}\right\rangle
\left\langle \theta _{j}\right\vert $ with $\left\vert \theta
_{1}\right\rangle =\cos \theta \left\vert 0\right\rangle +e^{i\phi
}\sin \theta \left\vert 1\right\rangle $ and $\left\vert \theta
_{2}\right\rangle =-\cos \theta \left\vert 1\right\rangle +e^{-i\phi
}\sin \theta \left\vert 0\right\rangle $ ($0\leq \theta \leq 2\pi $,
$0\leq \phi \leq 2\pi $). From (\ref{QI2}), it is clear that
different choices of $\left\{ \Pi _{j}^{B}\right\} $ may lead to
different values of $J_{q}\left( \rho _{A\colon B}\right) $. The
minimum difference between $I_{q}\left( \rho _{A\colon B}\right) $
and $J_{q}\left( \rho _{A\colon B}\right) $, called quantum discord
(QD) \cite{Ollivier}, is used to describe the quantum correlation of
a bipartite quantum system
\begin{equation}
D\left( \rho _{A\colon B}\right) =\min_{\left\{ \Pi _{j}^{B}\right\}
}\left[ I_{q}\left( \rho _{A\colon B}\right) -J_{q}\left( \rho
_{A\colon B}\right) \right]  \label{QD1}
\end{equation}%
or equivalently
\begin{equation}
D\left( \rho _{A\colon B}\right) =I_{q}\left( \rho _{A\colon B}\right)
-\max_{\left\{ \Pi _{j}^{B}\right\} }\left[ J_{q}\left( \rho _{A\colon
B}\right) \right] .  \label{QD2}
\end{equation}

From (\ref{TC}) and (\ref{QD2}), the classical correlation contained in a
quantum system is defined as \cite{Henderson}%
\begin{equation}
C\left( \rho _{AB}\right) \equiv I_{q}\left( \rho _{A\colon B}\right)
-D\left( \rho _{A\colon B}\right) =\max_{\left\{ \Pi _{j}^{B}\right\} }\left[
S\left( \rho _{A}\right) -S_{\left\{ \Pi _{j}^{B}\right\} }\left( \rho
_{A\mid B}\right) \right] .  \label{CC}
\end{equation}

In our investigation, entanglement is qualified by the Wootters
concurrence \cite{concurrence}. For given density matrix $\rho
_{AB}$ of a bipartite system $AB$ , the concurrence is defined as
$C=\max \left\{ 0,\sqrt{\lambda _{1}}-\sqrt{\lambda
_{2}}-\sqrt{\lambda _{3}}-\sqrt{\lambda _{4}}\right\} $, where
${\lambda _{i}}$ ($i=1,2,3,4$) are the eigenvalues of the matrix $
R=\rho _{AB}\left( \sigma _{A}^{y}\otimes \sigma _{B}^{y}\right)
\rho _{AB}^{\ast }\left( \sigma _{A}^{y}\otimes \sigma
_{B}^{y}\right)$, arranged in decreasing order of magnitude, $\rho
_{AB}^{\ast }$ is the complex conjugate of $\rho _{AB}$ and $\sigma
_{A,B}^{y}$ are the Pauli matrices for systems $A$ and $B$. The
concurrence attains its maximum value $1$ for maximally entangled
states and $0$ for separable states.

\section{Results and Discussion}

The master equation (\ref{master equation}) can be easily solved by the
fourth-order Runge-Kutta method in the representation spanned by the
eigenstates of $H_{S}$. We take the evolution time long enough such that the
final density matrix reaches steady state $\rho_{st}$. Then, according the
definitions on QD and concurrence given in the preceding section, we can
investigate the influence of the bath temperature, multi-spin interaction
and external magnetic field on the QD and concurrence of the spin chain. In
the calculation, we set the coupling constant $J=1$. It means that all the
interaction constants in the Hamiltonian are rescaled by the XX spin chain
coupling strength. We also assume that the decay rate is spectrum
independent, i.e. $\gamma (\omega )=\gamma $.

\begin{figure}[htbp]
\centering \includegraphics[width=12.0cm]{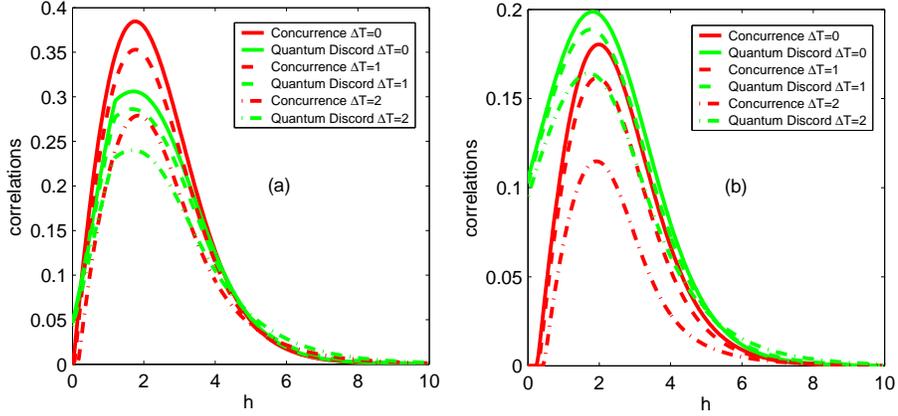}
\caption{Steady-state QD (green lines) and concurrence (red lines)
as a function of the field $h$ with various values of the
temperature difference. The other parameters are chosen to be
$\protect\gamma =0.01,T_{M}=1.8$ and $k=2$. The figures (a) and (b)
are for the spin pairs $13$ and $23$, respectively.}
\label{fig:Fig2}
\end{figure}

\begin{figure}[htbp]
\centering \includegraphics[width=12.0cm]{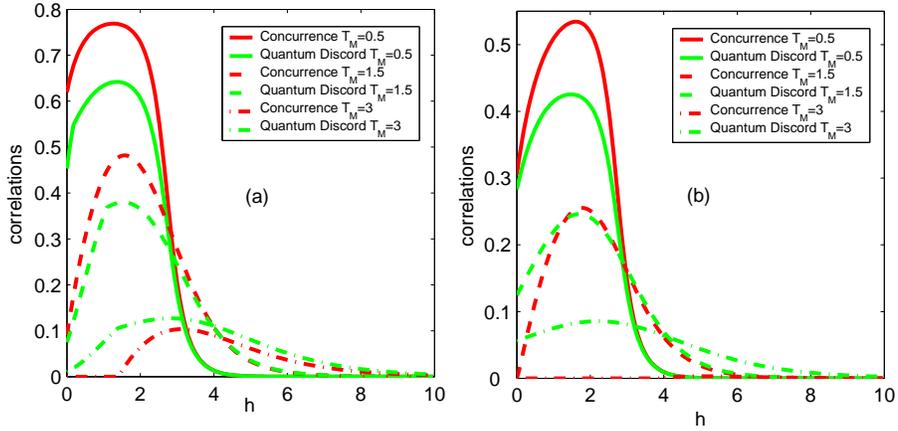}
\caption{Steady-state QD (green lines) and concurrence (red lines)
as a function of the field $h$ with various values of the mean
temperature. The other parameters are chosen to be $\protect\gamma
=0.01,\Delta T=0.5$ and $k=2$. The figures (a) and (b) are for the
spin pairs $13$ and $23$, respectively.} \label{fig:Fig3}
\end{figure}

In Figs. 2 and 3, the steady-state QD and concurrence of spin pairs
$13$ and $23$ are shown as a function of the magnetic field for
various values of the temperature difference $\Delta T=T_{1}-T_{3}$
and of the mean temperature $T_{M}=(T_{1}+T_{3})/2$. In these
figures, we see that both the QD and concurrence first increase with
increasing of the field, get maximal values and then decay to zero.
As either the temperature difference or the mean temperature
increases, in general, the QD and concurrence are diminished.
Comparing Fig. 2 with Fig. 3, we notice that the mean temperature
affects more strongly the concurrence than the temperature
difference. The sudden death of concurrence as shown in Fig. 3
indicates that the QD is more robust against the mean temperature
than the concurrence. From these figures, we come to the conclusion
that the QD and concurrence can be enhanced by switching on the
properly large magnetic field if both the mean temperature and
temperature difference are not large.

\begin{figure}[htbp]
\centering \includegraphics[width=12.0cm]{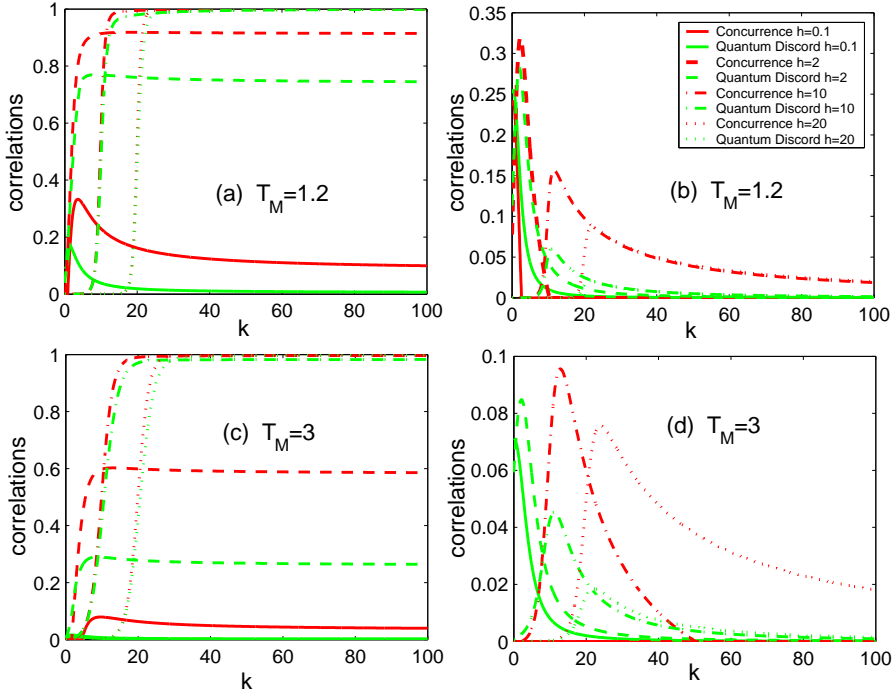}
\caption{Steady-state QD (green lines) and concurrence (red lines) as a
function of the three-spin interaction strength $k$ with various values of
the mean temperature and of the field strength. The figures (a) and (c), and
(b) and (d) are for spin pairs $13$ and 23, respectively. The other
parameters are chosen to be $\protect\gamma =0.01$ and $\Delta T=0.8$. The
symbols shown in the inset of Fig. 4(b) are applicable to all the curves of
Fig. 4.}
\label{fig:Fig4}
\end{figure}

\begin{figure}[htbp]
\centering \includegraphics[width=12.0cm]{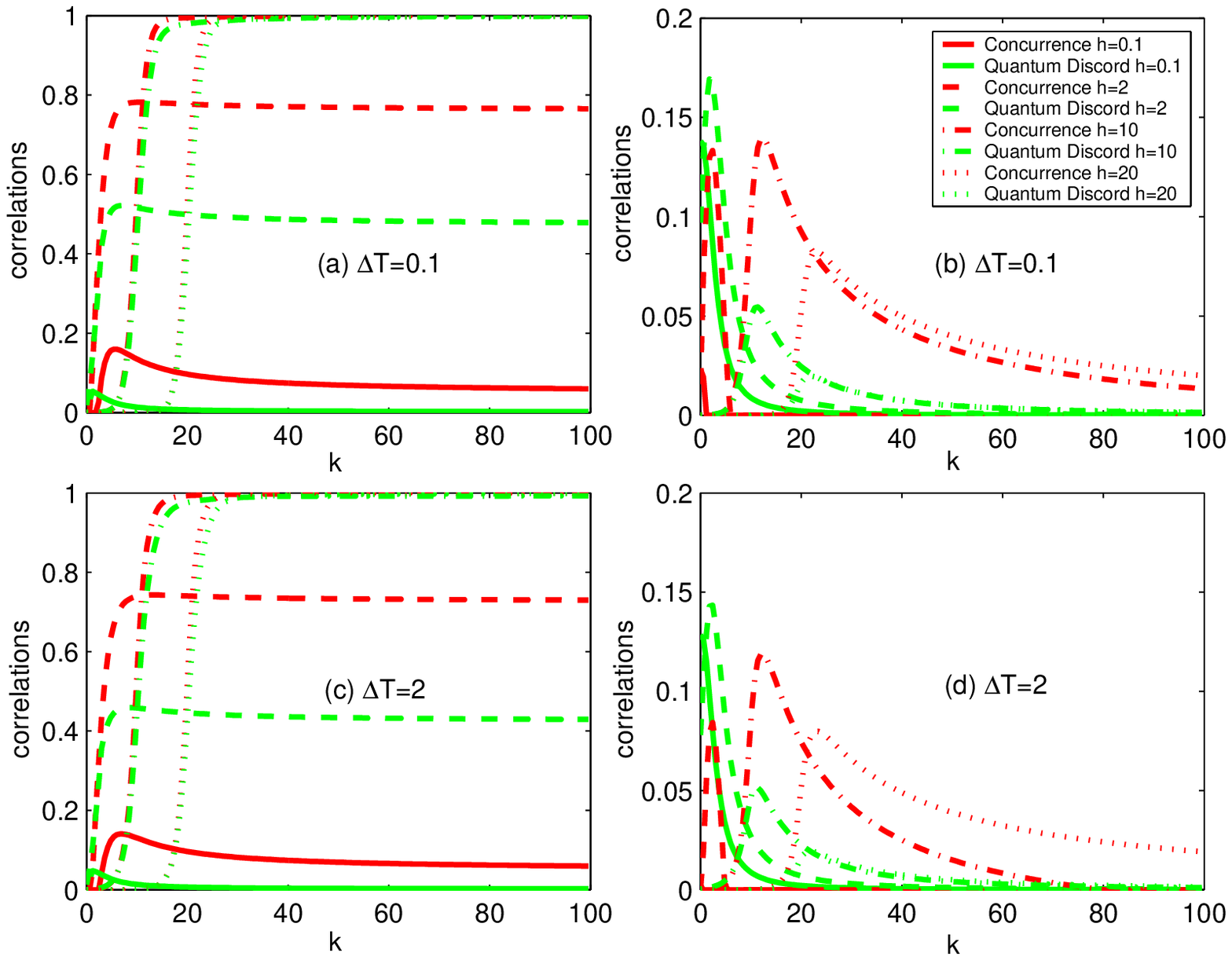}
\caption{Steady-state QD (green lines) and concurrence (red lines) as a
function of the three-spin interaction strength $k$ with various values of
the temperature difference and of the field strength. The figures (a) and
(c), and (b) and (d) are for spin pairs $13$ and 23, respectively. The other
parameters are chosen to be $\protect\gamma =0.01$ and $T_{M}=2$. The
symbols shown in the inset of Fig. 5(b) are applicable to all the curves of
Fig. 5.}
\label{fig:Fig5}
\end{figure}

\begin{figure}[htbp]
\centering \includegraphics[width=12.0cm]{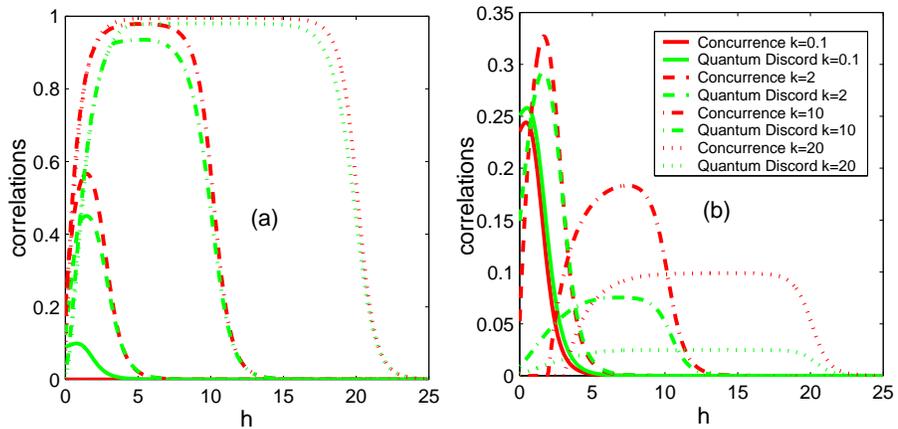}
\caption{Steady-state QD (green lines) and concurrence (red lines)
as a function of the magnetic field strength $h$. The figures (a)
and (b) are for spin pairs $13$ and $23$, respectively. The
parameters are chosen to be $\protect\gamma =0.01,T_{M}=1.2$ and
$\Delta T=0.8$. The symbols shown in the inset of Fig. 6(b) are
applicable to the curves of Fig. 6 (a).} \label{fig:Fig6}
\end{figure}

Figures 4 and 5 show the steady-state QD and concurrence of spin pairs $13$
and $23$ as a function of the three-spin interaction strength $k$ for
various values of the magnetic field strength. In these figures, we see that
the QD and concurrence for spin pair $13$ first increase with the three-spin
interaction and then get a plateau. It is very interest that the plateau can
be raised to the maximum level by increasing the magnetic field strength. As
shown in Figs. 4 and 5, this feature can be maintained even if the mean
temperature is high and the temperature difference is large. As for the spin
pair $23$, figures 4 and 5 show that its QD and concurrence first increase
with the three-spin interaction, get peaks and then decay to zero. In Fig.
6, the steady-state QD and concurrence of spin pairs $13$ and $23$ as a
function of the magnetic field strength $h$ with various values of the
three-spin interaction strength $k$. In Fig. 6(a), we observe that the QD
and concurrence of spin pair $13$ can get the maximum level plateau by
increasing the field if the three-spin interaction is enough strong. The
maximum plateau width is enlarged as the interaction strength increases.
Thus, we can maintain the extreme QD and concurrence of the spin pair $13$
by holding the strong interaction and magnetic field.

\begin{figure}[htbp]
\centering \includegraphics[width=12.0cm]{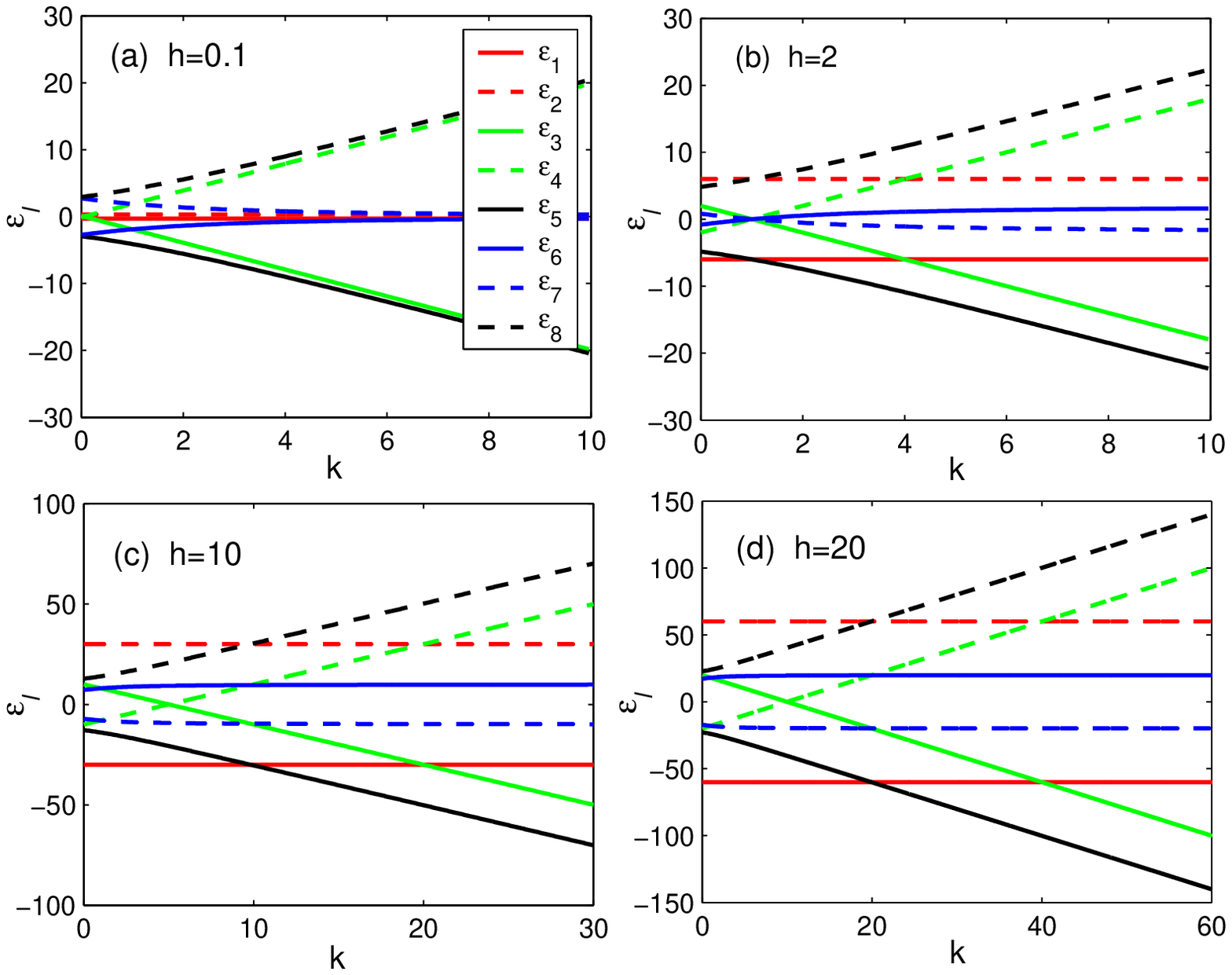}
\caption{Eigenenergy $\protect\varepsilon_l$ of the Hamiltonian (1) as a
function of the three-spin interaction strength $k$ with various values of
the field strength. The symbols shown in the subset of Fig. 7(a) are
applicable to all the curves in the figures.}
\label{fig:Fig7}
\end{figure}

\begin{figure}[tbph]
\centering \includegraphics[width=12.0cm]{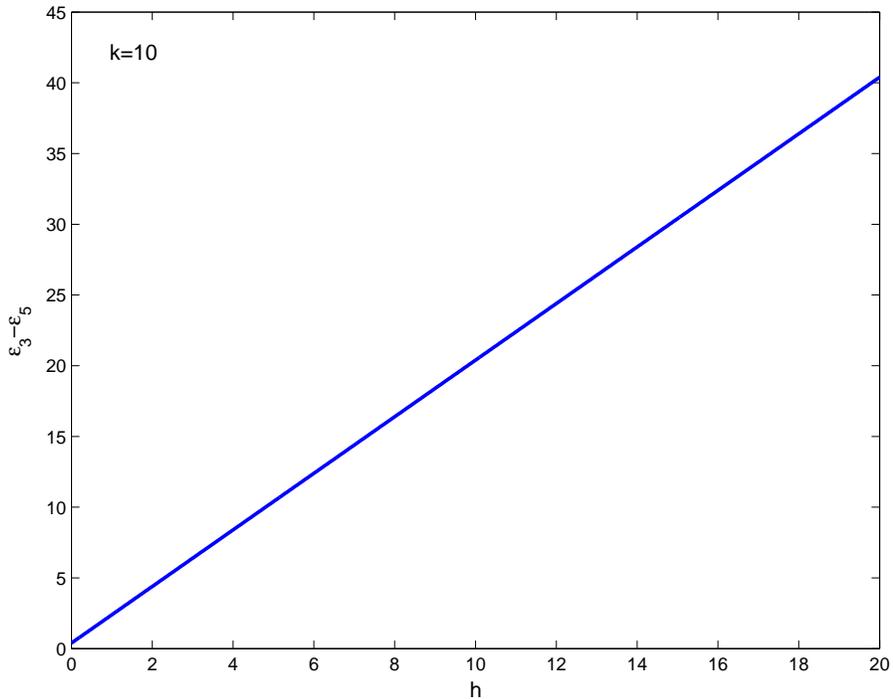}
\caption{The energy difference between $\protect\varepsilon _{3}$ and $%
\protect\varepsilon _{5}$ as a function of the field with a fixed value of
the interaction, $k=10$.}
\label{fig:Fig8}
\end{figure}

In order to find out the physical reasons for the observed
phenomena, we first analyze the eigenvalues $\varepsilon _{l}$
($l=1,...,8$) of the Hamiltonian (1) as a function of the
interaction and magnetic field strengths. When the field is
weak($h<<k$), the eigenstate $\vert\phi_5\rangle $ is the ground
state of the spin chain since the eigenenergy $\varepsilon
_{5}=-h-k-\sqrt{8+k^2}$ is the smallest one as shown in Fig. 7(a).
It is noted that $\varepsilon _{1}=-3h$. Thus, the eigenstate
$\vert\phi_1\rangle$ becomes the ground state when $h>k$. The two
states have the energy crossing around the point $h=k$ as shown in
Figs. 7 (b)-7(d). Since then, the eigenstate $\vert\phi_5\rangle$
becomes the ground state. As the interaction strength $k$ further
increases, the state $\vert\phi_3\rangle$ with the eigenenergy
$\varepsilon _{3}=-h-2k$ crosses with the state $\vert\phi_1\rangle$
and becomes the first excited state of the spin chain. In Fig. 8,
the energy difference between the eigenstates $\vert\phi_5\rangle$
and $\vert\phi_3\rangle$ is plotted as a function of the magnetic
field. We see that the energy splitting linearly increases as the
magnetic field increases. In fact, we have $\varepsilon
_{3}-\varepsilon _{5}\approx 2h$ when $k$ is large.

\begin{figure}[tbph]
\centering \includegraphics[width=12.0cm]{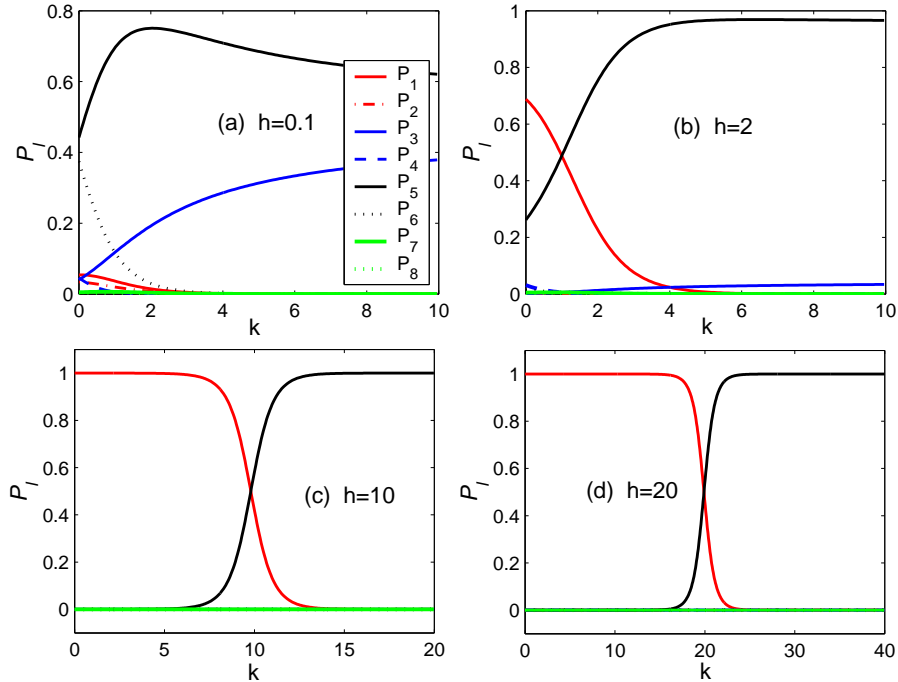}
\caption{Eigenstate occupation probabilities of the steady-state
density matrix $\protect\rho_{st}$ as a function of the three-spin
interaction strength $k$ with various strength of the field. The
other parameters are chosen to be $\protect\gamma =0.01,T_{M}=1.2$
and $\Delta T=0.8$. The symbols shown in the subset of Fig. 9(a) are
applicable to all the curves of Figs. 9.} \label{fig:Fig9}
\end{figure}

Figure 9 shows the eigenstate occupation probabilities which are
defined as $P_l=tr(\vert\phi_l\rangle\langle\phi_l\vert\rho_{st})
=\langle\phi_l\vert\rho_{st}\vert\phi_l\rangle$ as a function of the
three-spin interaction. We see that $\vert\phi_1\rangle$ is the most
populated state when $k<<h$, $\vert\phi_1\rangle$ and
$\vert\phi_5\rangle$ cross and take the same probability around the
point $k=h$ when $h$ is large, and then $\vert\phi_5\rangle$ becomes
the most populated state and takes over all the occupation
probability, as shown in Figs. 9(b)-9(d).

\begin{figure}[tbph]
\centering \includegraphics[width=12.0cm]{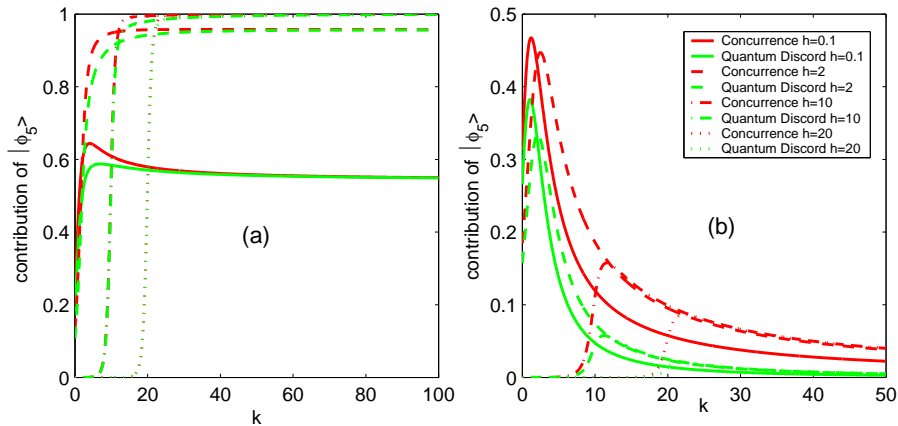} \caption{The QD
and concurrence of $\left\vert \protect\phi_{5}\right\rangle $ as a
function of the three-spin interaction with various values of the
field. The other parameters are chosen to be $\protect\gamma
=0.01,T_{M}=1.2 $ and $\Delta T=0.8$. The curves (a) and (b)
correspond to correlations of spin pairs $13$ and $23$,
respectively. The symbols shown in the subset of Fig. 10(b) are also
applicable to all the curves of Fig. 10(a).} \label{fig:Fig10}
\end{figure}

Therefore, when the magnetic field and three-spin interaction are
strong, the possible thermal excited transition is one from $|\phi
_{5}\rangle $ to $|\phi _{3}\rangle $, which is induced by the
thermal resources interacting with the spins $1$ and $3$,
respectively. However, $\varepsilon _{3}-\varepsilon _{5}\approx
2h$, as discussed above. Thus, the thermal excitation can be mostly
depressed if $h\gg T_{1},T_{3}$. In this case, the spin chain is
nearly decoupled from the thermal resources. Thus, the QD and
concurrence of the spin system are determined by the most populated
eigenstate, i.e. the ground state of the spin system. These results
mean that the eigenstate $|\phi _{5}\rangle $ makes the most
contribution to the QD and concurrence of the spin pairs $13$ and
$23$ when the magnetic field and three-spin interaction are strong
enough. In Fig. 10, the QD and concurrence of $|\phi _{5}\rangle$
are plotted as a function of the three-spin interaction $k$.
Comparing Fig. 10 with Fig. 4, we see that the above conclusion is
really true.

\section{Summary}

We investigate the quantum discord and concurrence of the three-spin
chain which ends are coupled to two thermal reservoirs at different
temperatures. Besides the XX-type nearest-neighbor two-spin
interaction a three-spin interaction is also included and a
homogenous magnetic field is applied to each spin. For fixed
temperatures of the thermal reservoirs, we find that the extreme
steady-state QD and concurrence of the two end spins can always be
created by raising the magnetic field strength with a strong
multi-spin interaction. We show that the energy gap between the most
populated ground state and the first excited state of the spin chain
can become much larger than the thermal excitation energy when the
magnetic field and multi-spin interaction are strong enough. In this
way, the thermal excitation induced by the thermal reservoirs is
nearly depressed and the spin chain is decoupled from the thermal
environments. As a result, the QD and concurrence of the spin chain
are totaly determined by the most populated ground state of the spin
chain. The present results may provide a useful approach to control
coupling between a quantum system and its environment.

\section*{Acknowledgments}

This work was supported by the National Basic Research Program of China (973
Program) No. 2010CB923102, Special Prophase Project on the National Basic
Research Program of China (Grant No.2011CB311807) and the National Nature
Science Foundation of China (Grant No. 11074199).

\end{document}